\documentstyle[twoside,fleqn,espcrc2,amsmath,cite,epsfig]{article}

\def\cO#1{{\cal{O}}\left(#1\right)}

\def\as{\alpha_{\mbox{\scriptsize s}}}

\def\be{\beta}

\def\out{\mbox{\scriptsize out}}

\def\Ko{K_{\out}}

\def\ee{e^+e^-}

\def\conf{\delta}

\def\cR{{\cal{R}}}

\def\cF{{\cal{F}}}

\def\abs#1{\left| \: #1 \: \right|}%

\def\bmu{\bar{\mu}}
\def\bnu{\bar{\nu}}
\def\bKo{\bar{K}_{\out}}
\def\LQCD{\Lambda_{\mbox\scriptsize QCD}}

\def\vk{\vec{k}}

\def\bit{\begin{itemize}}
\def\eit{\end{itemize}}

\newcommand\app[3]{{{\it Acta Phys. Polon }{\bf B #1} (#2) #3}}

\newcommand\cpc[3]{{{\it Comput. Phys. Commun. }{\bf #1} (#2) #3}}

\newcommand\jhep[3]{{{\it JHEP }{\bf #1} (#2) #3}}

\newcommand\npb[3]{{{\it Nucl. Phys. }{\bf B #1} (#2) #3}}
\newcommand\npps[3]{{{\it Nucl. Phys. }{\bf #1} {\it(Proc. Suppl.)} (#2) #3}}

\newcommand\plb[3]{{{\it Phys. Lett. }{\bf B #1} (#2) #3}}

\newcommand\prep[3]{{{\it Phys. Rep. }{\bf #1} (#2) #3}}

\newcommand\zpc[3]{{{\it Z. Physik }{\bf C #1} (#2) #3}}

\newcommand\epjdirc[3]{{{\it Eur. Phys. J. Direct }{\bf C #1} (#2) #3}}

\title{Perturbative QCD analysis of near-to-planar three-jet events.\thanks{%
    Talk presented by AB at QCD 00 Euroconference, Montpellier,
    July 2000.}}

\author{
\vspace{-3.5cm}
\begin{flushright}
  Bicocca--FT--00--14\\
  Pavia--FNT/T--00--17\\
  hep-ph/0010052
\end{flushright}
\vspace{2.0cm}
    A. Banfi\address{Dipartimento di Fisica, Universit\`a di Milano, and 
    INFN, Sezione di Milano, 20133 Milano,
    Italy} and G. Zanderighi\address{Dipartimento di Fisica Nucleare e
    Teorica, Universit\`a di Pavia, and INFN Sezione di Pavia, 27100
    Pavia, Italy}}

\begin{document}

\begin{abstract}
  We present the all-order resummed $\Ko$ distribution as a measure of 
  aplanarity in three-jet events.
 
\end{abstract}

\maketitle

\section{Introduction}

In recent years much theoretical and experimental effort has been devoted 
to the study of the so-called jet-shape variables. 
Such observables are useful to describe the spatial distribution 
of final state particles in high
 energy 
hadronic processes, like  $e^+e^-$ annihilation and DIS.
For many of these (e.g. thrust and heavy-jet mass \cite{thrust},
 C-parameter \cite{Cpar} and the
 jet Broadenings \cite{broad})
perturbative (PT) distributions are already available, and it is even possible 
to estimate the power-suppressed corrections due to 
confinement effects (for reviews see\cite{1/Q},\cite{Blois}).  

For two-jet configurations,  the  
kinematical region in 
which the two jets are ``pencil-like'' turns out to be particularly interesting
, since here multiple soft gluon 
radiation effects become essential. 
We meet an analogue of this situation for three-jet events if we study  
near-to-planar 
configurations. The variable we choose as a measure of event aplanarity 
is $\Ko$,
defined
 as the 
sum of the moduli of the momenta lying out of the event plane.

The aim of this paper is to present the main features of $\Ko$ perturbative
distribution, while all computational details may be found in \cite{noi}. 

The calculation is performed at the ``state of art'' level, which means
  all-order resummation of double- (DL) and single-logarithmic (SL)
  contributions due to soft and collinear gluon radiation effects,
  two-loop analysis of gluon radiation probability and
  matching the resummed logarithmic expressions with the exact
  $\cO{\as^2}$ results.

In Section 2 basic definitions and notations are introduced, together with a
 discussion
on the sources of SL corrections. These are examined in detail in Section 3
, which 
contains the complete expression for $\Ko$ distribution to SL accuracy.
 Gluon radiation 
effects
 are included in the so-called ``radiator'', which is described in
 Section 4. 
In Section 5  is shown the final answer for the distribution. We finally
 draw some conclusions and present further developments of our 
research (Section 6).

\section{Definition of the observable}
\label{sec:def}
As it is known, a three-jet event in $\ee$ annihilation is generated by a
quark-antiquark pair which radiates a hard gluon.
For kinematical reasons these three partons lie in a
plane. 
When additional partons are emitted this is no longer true,
but it is still possible to define an event-plane and to measure the
radiation out of this plane.

First we introduce the thrust and thrust-major of the event ($Q$ is the
 $\ee$ centre-of-mass energy),
\begin{equation}
  \label{eq:thrust}
\hspace{-0.8cm}
  T\,Q = \max_{\vec{n}} \left\{ \sum_{h} \abs{\vec{n}\vec{p}_h}
    \right\} \>=\>
\sum_{h} \abs{{p}_{hz}} \>,
\end{equation}
\begin{equation}
  \label{eq:thrustM}
\hspace{-0.8cm}
  T_M\,Q = \max_{\vec{n}\vec{n}_T=0} 
   \left\{ \sum_{h} \abs{\vec{n}\vec{p}_h}
    \right\} \>=\>  \sum_{h} \abs{{p}_{hy}} \>.
\end{equation}
We choose to fix the $z$ and $y$ axis along the thrust and thrust-major axis 
respectively.
Orientation of these two axes is determined by imposing that 
the momentum of the most energetic particle has a positive
$z$-component, while the momentum of the second most 
energetic particle has a positive $y$
component.
 Hereafter we attribute $p_h$ to the {\bf right}
hemisphere ({\bf left} hemisphere ) if $p_{hz}>0$
($p_{hz}<0$). Similarly $p_h$ is in the {\bf up}  hemisphere 
({\bf down} hemisphere) if $p_{hy}>0$ ($p_{hy}<0$). 
The $T$-$T_M$ plane is called the event plane, and the sum of the moduli of the
 out-of-plane
momenta gives the thrust-minor:
\begin{equation}
  \label{eq:thrustm}
\hspace{-0.8cm}
  T_m\,Q = \sum_{h} \abs{{p}_{hx}} \>\equiv\> \Ko \>.
\end{equation}
Such definitions, together with energy-mo\-men\-tum conservation, force the 
transverse momentum  with respect to the thrust axis $k_t\equiv (k_x,k_y)$
 to be conserved 
separately in the right and left hemisphere. Similarly the $x$-component of
$k_t$ is conserved independently in the up and down hemisphere.

Near-to-planar three-jet events belong to the kinematical region 
$T\sim T_{M} \gg T_{m}$. At the parton level such events can be treated as a
 hard quark-antiquark-gluon system accompanied by soft and collinear partons.

At the Born level no radiation is present, the three hard partons are in a 
plane
and, as a consequence, $\Ko$ trivially vanishes.

\begin{figure}[tb]
  \begin{center}
   \epsfig{file=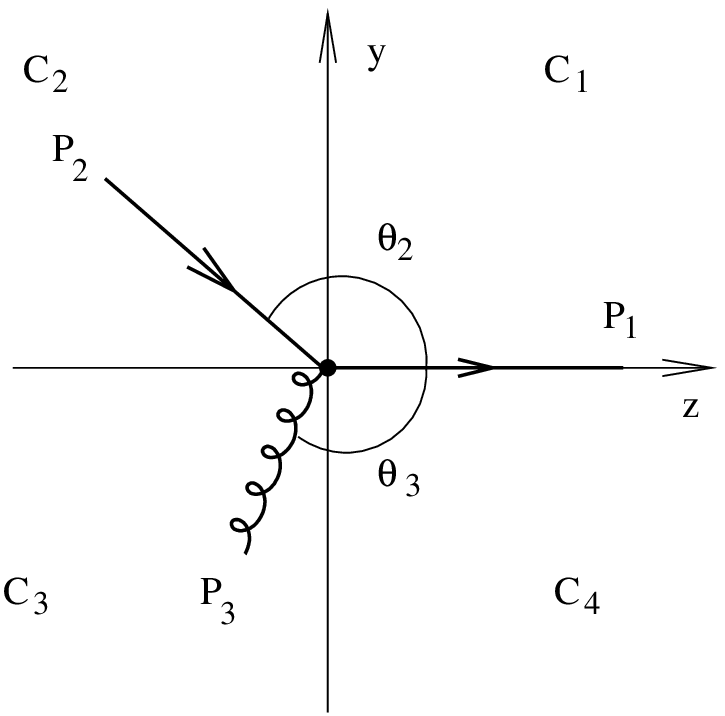,width=0.30\textwidth}
    \caption{The  Born
      configuration with $\conf=3$ for $T=0.75$ and
      $T_M=0.48$ .  The
      thrust $T$ and thrust major $T_M$ are along the $z$- and $y$-axis
      respectively. The four regions $C_{\ell}$ in the phase space are
      indicated.}
    \label{fig:conf}
  \end{center}
\end{figure}

Hereafter we denote by  $p_1,p_2$ 
and $p_3$ the energy ordered ($p_{10}>p_{20}>p_{30}$) hard parton momenta.
There are three possible Born configurations, according whether
the momentum of the gluon is  $p_1,p_2$ or $p_3$. They are named by labelling
a configuration with an index $\conf$ which is set equal to the gluon momentum
 index. In order to clarify notations, Figure~\ref{fig:conf} shows the 
configuration 
corresponding to $\conf =3$. 

Secondary parton radiation bring particle momenta out of the event plane, 
so that $\Ko$ no
 longer vanish. We therefore study the ``integrated'' $\Ko$ distribution
$\Sigma$, defined by
\begin{equation}
  \label{eq:DLdist}
\hspace{-0.8cm}
 \frac{d\sigma_{\conf}}{dT\>dT_M\>d\Ko}\equiv C(\as)\>M_{\conf}^2(T,T_M)\>
 \frac{d\Sigma_{\conf}(\Ko)}{d\Ko}\>.
\end{equation}
Here $M_{\conf}^2(T,T_M)$ denotes the hard matrix element for the production of
 a quark-antiquark-gluon ensemble in the configuration $\conf$ for given values
of $T$ and $T_M$. 
The coefficient function $C(\as)$, which has an
expansion in powers of $\as(Q)$, takes into account all the effects 
coming from the hard phase space region of emitted partons.     

At the DL level, only  gluons which are both soft and collinear
 can contribute to $\Sigma$. An all-order resummation of such contributions 
gives rise to the Sudakov exponent
\begin{equation}
\hspace{-0.8cm}
\Sigma(\Ko)\simeq \exp 
   \left\{-(2C_F+C_A)\frac{\as}{\pi}\ln^2\frac{\Ko}{Q}\right\},
\end{equation}
which is the product of three independent factors, one for each
hard radiating parton.
This simple structure reflects the fact that a gluon which is collinear to
 one of the three hard partons
is not influenced by the presence of the remaining two hard emitters.
Only soft and non-collinear gluons are able to explore the inter-jet region 
and see
the topological structure of the event. 
These interference effects are typical of three-jet events, and give rise to
SL corrections to the above distribution.  
The other sources of SL contributions are present also in two-jet
 shape 
variables. They can be obtained by properly treating 
 hard parton recoil, by taking into account hard collinear parton splitting and
by determining the argument of the running coupling.

\section{SL corrections}
\label{sec:SL}
In this section we describe in detail how the SL contributions 
discussed in the previous section modify $\Ko$ distribution. 

Hard collinear parton decay is taken into account by simply 
replacing the soft matrix element with the complete Altarelli-Parisi splitting
functions.

A two loop analysis of soft gluon radiation \cite{Catani} is needed to 
determine the 
argument of the running coupling. The result is that, following \cite{DLMS},
in the physical scheme \cite{CMW}, for each emitting dipole
$\as$ depends on the invariant transverse momentum of the emitted gluon with 
respect to the radiating dipole.
Furthermore, after the $k_y$ integration,
one can safely replace the argument by twice the $k_x$-component,
which is dipole independent.

Exponentiation of
hard parton recoil effects is 
possible introducing small recoil momenta $q_a$
\begin{equation}
  \label{eq:rec}
\hspace{-0.8cm}
p_a=P_a+q_a\qquad a=1,2,3
\end{equation}
This allows to split parton phase space into a hard and a soft part.
The hard part is embedded in the hard matrix element, while the soft 
part can be exponentiated together with soft gluon emission probability
 via a multiple 
Fourier-Mellin transform. The integrated $\Ko$ distribution assumes now 
the form:
\begin{equation}
  \label{eq:SLdist}
\hspace{-0.8cm}
\begin{split}
\Sigma(\Ko)&=\int\frac{d\nu}{2\pi i\nu}e^{\nu\Ko}\sigma(\nu)\\
\sigma(\nu)&=\int[d\gamma d\beta]I(\beta,\gamma)e^{-\cR (\nu,\beta,\gamma)}\>,
\end{split}
\end{equation}
where $\gamma$ and  $\beta_a$ are the Fourier variables conjugated to
$q_{1y}$ and $q_{ax}$ respectively.

All the effects of gluon radiation are contained in the ``radiator'' $\cR$: 
\begin{equation}
  \label{eq:rad}
\hspace{-0.8cm}
\cR(\nu,\beta,\gamma)=\int\frac{d^3\vk}{\pi\omega}w(k)
                      \left(1-\sum_{\ell=1}^4 u_\ell(k)\right).
\end{equation}
This function is given by the two-loop one gluon emission probability $w(k)$
($k\equiv(\omega,\vk)$ is the gluon four-momentum), multiplied by a 
combination of the ``sources'' $u_\ell(k)$ and integrated over the gluon phase
 space.  
The  sources $u_\ell(k)$  indicate how particle momenta contribute to $\Ko$.
They are 
one for each quadrant, since hard
parton recoil effects are different in each of this four phase space regions.  
Explicit expressions for $w(k)$, for the function $I(\beta,\gamma)$ and for 
the sources $u_\ell(k)$ may be found in \cite{noi}. 

We stress that, due to the presence of soft inter-jet gluons, we expect the
 radiator to be a
function of $T$ and $T_M$.

\section{The PT radiator}
\label{sec:rad}

 The relevant region for the perturbative contribution to the radiator is
$\nu\gg 1$. 
In this limit, for a generic configuration $\conf$, equation 
 \eqref{eq:rad} becomes  
\begin{equation}
\label{eq:RadTT}
\hspace{-0.8cm}
\begin{split} 
\cR_{\conf} &= C_2^{(\conf)}\>r(\bmu_2,Q_2^2)+C_3^{(\conf)}\>r(\bmu_3,Q_3^2)\\
              &+ \frac{C_1^{(\conf)}}{\pi}\int_0^\infty \!\!\frac{dy}{1+y^2}
\>\left[\,r(\bmu_{12},Q_1^2) \!+\! r(\bmu_{13},Q_1^2)\,\right],
\end{split}
\end{equation}
where  $r(\bmu,Q^2)$ is the DL function
\begin{equation}
  \label{eq:r}
\hspace{-0.8cm}
  r(\bmu,Q^2)=\int_{1/\bmu}^Q\frac{dk_x}{k_x}\frac{\as(2k_x)}{\pi}
\ln\frac{Q^2}{k^2_x}\>.
\end{equation}

It turns out that even to SL accuracy 
the radiator is the sum of three ``independent'' contributions, each one 
corresponding to one of the three hard emitting parton and proportional to 
 its colour charge $C_a^{(\conf)}$, which equals $C_F$ or $C_A$ depending
whether the
 parton $a$ is 
a quark or a gluon.

The three momentum scales $Q_a$ are given by
\begin{equation}
\label{eq:scalesT}
\hspace{-0.8cm}
\begin{split}
Q_{a}^2  & =  \frac{p_{ta}^2}{4}e^{-g_a}\>, \\
p_{ta}^2 & =  2\>\frac{(P_bP_a)(P_aP_c)}{(P_bP_c)}\>.
\end{split}
\end{equation}
They depend on the event geometry and, in particular,
 each one is proportional to
the invariant transverse momentum of the emitting parton $a$
with respect to the  
other two. This effect is due to the presence of soft, but non-collinear 
gluons, whose radiation is sensitive to the topology of the event.
The factor $g_a$ comes from the hard part of the splitting functions:  its
value is $3/2$ if parton $a$ is a quark and $\be_0 / 2N_c$ if 
parton $a$ is a gluon.
 
Hard parton recoil affects the radiator through the scales $\bmu$ 
(see \cite{noi} for their definition).

\section{$\Ko$ distribution}
\label{sec:kodist}
Using the following simplified notations 
\begin{equation}
\hspace{-0.8cm}
\begin{split}
&\bKo \>= e^{-\gamma_E}\Ko\>,\\
&R_{\conf}(\bnu)\> =\sum_{a=1}^3 C_a^{(\conf)}r(\bnu,Q_a)\\
& r'(\bnu)=\frac{\as(
k_x)}{\pi}\ln \frac{Q^2}{k^2_x}\>,\qquad
k_x\equiv 1/\bnu\>,\\
&R'(\bnu)=(2C_F+C_A)r'(\bnu)
\end{split}
\end{equation}
we can give the final result for $\Ko$ distribution:
\begin{equation}
  \label{eq:distfin}
\hspace{-0.8cm}
   \Sigma_{\conf}(\Ko) = e^{-R_{\conf}\left(\bKo^{-1}\right)} \cdot 
\frac{\cF_{\conf}\left({\bKo}^{-1}\right)}
{\Gamma\left(1+R'(\bKo^{-1})\right)}\>,
\end{equation}

The SL function $\cF_{\conf}$ contains the contribution of hard parton
 recoil. At first order in $r'$ it turns out to be
\begin{equation}
\label{cf1}
\hspace{-0.8cm}
\cF_{\conf}= 
 1-\ln 2 \left(2\, C_1^{(\conf)}+C_2^{(\conf)}+C_3^{(\conf)}\right)
\cdot r'\>.
\end{equation}
It may seem surprising  that the contribution from the parton lying on the
 thrust axis gets doubled with respect to the other two, but there is a simple
argument for this, which exploits momentum conservation. 
 When a soft gluon is emitted from parton 2 (or 3) 
both the gluon transverse momentum and the hard parton recoil momentum 
lie in the up(down)-region, and momentum conservation is
possible. 
On the contrary, when a gluon is emitted from parton 1, if its $k_t$ 
lies in the up-region, the recoil momentum lies in the down region.
Because $k_x$ is conserved separately in
the up and down region, there has to be a recoil contribution from the
two remaining partons, which explains the result in \eqref{cf1}. 
This prediction has been found in agreement with numerical
simulations obtained with the Monte-Carlo program  EVENT2~\cite{event},  as 
shown in Figure~2.
\begin{figure}[tb]
  \begin{center}
   \epsfig{file=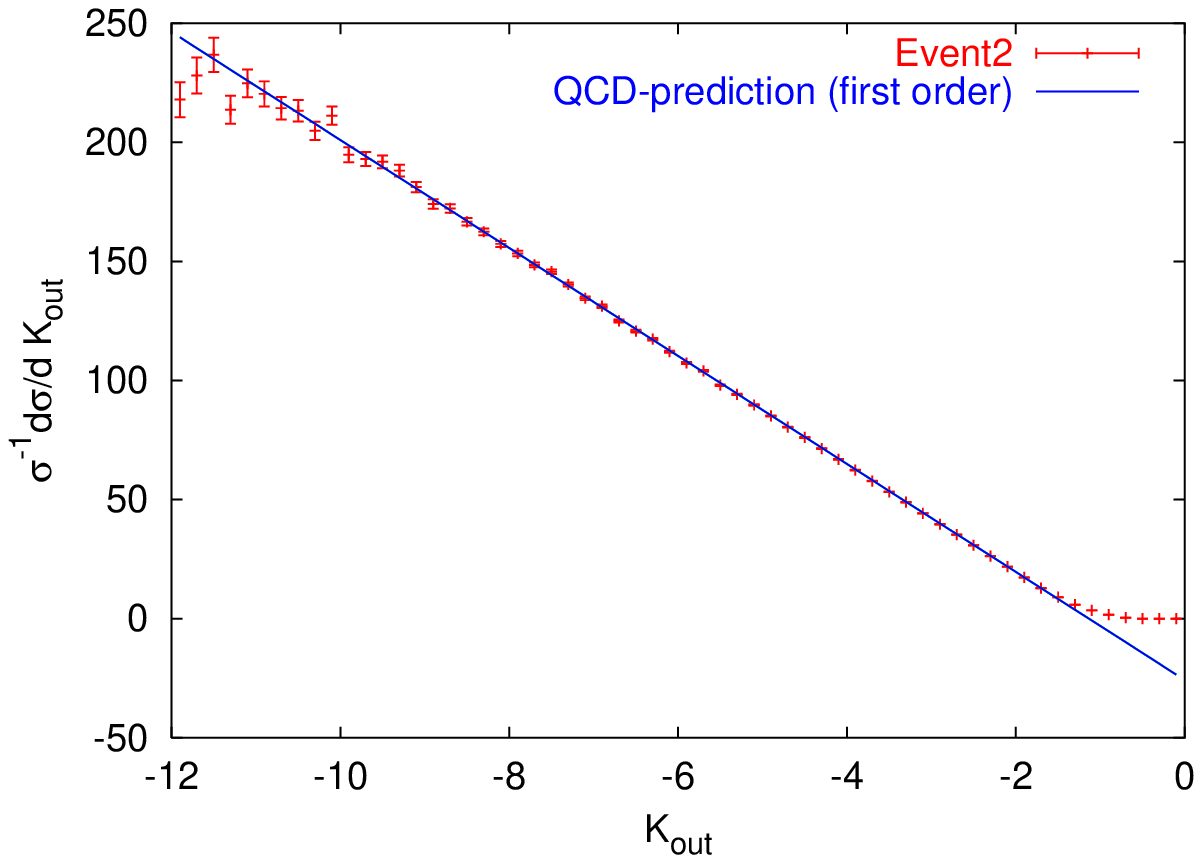,width=0.45\textwidth}
    \caption{Comparison of fixed order QCD
    distribution with numerical simulations based on the exact four parton
    matrix element for $y_3>0.1$ in units of $(\as/(2\pi))^2$.}  
      \end{center}
\label{fig:ev2}
\end{figure}

All effects of soft inter-jet  gluon radiation and of hard parton 
collinear splitting are embodied in the scales $Q_a$, which depend on 
$T$ and $T_M$.

\section{Conclusions and outlook}
The most interesting feature of $\Ko$ distribution in 
\eqref{eq:distfin} is
its geometry (i.e. $T$ and $T_M$) dependence, which allows
 to measure the effects of large 
angle gluon radiation.

In order to compare theory with experiments, one has to sum over all
possible configurations and to integrate over $T$ and $T_M$ varying in an 
appropriate range. At this point some problems may arise.

Till now the available data explore all values of $T$ and $T_M$, 
but the dominant $T\!\sim\! 1$ region has to be excluded for several reasons.
First of all in this region our 
calculation is incomplete because the hard matrix element becomes IR singular
 as the emitting gluon becomes soft or collinear to the
 quark-antiquark pair, so that an additional resummation should be performed.
Furthermore, in this region the definition of an event plane becomes
problematic, if not even meaningless.
Last but not least, even if we were able to perform the calculation,
the result would not be interesting, since we lose the richness of three-jet
event topology.
In order to avoid this problem one should look for data in a ``restricted''
 region, e.g.  $T\! <\! 0.9$. From an experimental point of view it is
more convenient to put a cut in the three-jet resolution parameter.
 If one uses the 
Durham recombination algorithm \cite{Durham}, a suitable choice is, for 
example, $y_3 >0.1$.

However, before any comparison with data may be attempted,
some theoretical work 
has still to be made.

 First of all one has to compare the resummed result 
with numerical NLO computation based on the exact two loop matrix element
 in order to check resummation to $\cO{\as^2\ln^2}$ and to
 to compute the coefficient function $C(\as)$ with the best achievable
accuracy. 

Then one has to match the resummed result with these fixed order
calculations to obtain an improved prediction in the whole phase space
considered. 

But indeed the greatest effort shall be made to compute hadronization effects,
the so-called non-perturbative (NP) ``power-corrections'', 
which are due to the emission of gluons 
with $k_t\sim\LQCD$.
We believe that these correction are of $1/Q$ type, enhanced by a factor 
$\ln \Ko/Q$, in analogy with what happens for the jet Broadenings
~\cite{broadNP}. 
  This will be the aim of our research in the forthcoming period, and we hope
that a complete (PT and NP) resummed $\Ko$ distribution may appear soon.

\paragraph{Acknowledgements}     

We are grateful to  Yuri Dokshitzer and Pino
Marchesini, since without their aid this work would had never been possible,
and to Gavin Salam for illuminating discussions and suggestions. 
\vspace{+2cm}

\end{document}